\documentclass[journal]{IEEEtran}

\usepackage[utf8]{inputenc}
\usepackage[T1]{fontenc}
\usepackage{amsmath}
\usepackage{amsfonts}
\usepackage{amssymb}
\usepackage{graphicx}
\graphicspath{{./}{img/}{figures/}}
\usepackage{cite}
\usepackage{url}
\usepackage{booktabs}
\usepackage{adjustbox}
\usepackage{caption}
\usepackage{subcaption}
\usepackage[hidelinks]{hyperref}

\numberwithin{equation}{section}
\renewcommand{\theequation}{\arabic{section}.\arabic{equation}}

\title{IsoNet: Face-Guided Speech Extraction for Compact Microphone Arrays}

\author{Dinanath Padhya,
Sajen Maharjan,
Binita Adhikari,
and~Ishwor Raj Pokharel%
\thanks{Dinanath Padhya, Sajen Maharjan, Binita Adhikari, and Ishwor Raj Pokharel are with the Department of Electronics and Computer Engineering, Thapathali Campus, Institute of Engineering, Tribhuvan University, Kathmandu, Nepal (e-mail: dinanath@tcioe.edu.np).}}

\begin{document}

\maketitle

\begin{abstract}
    Listening to one chosen talker in a crowded room is effortless for people but hard for small devices, whose microphone arrays are too compact for classical beamforming to separate nearby voices. We present IsoNet, a select-to-listen system: the user selects a face in the camera view, and a 4-microphone array with a 9.4~cm aperture returns that person's speech. IsoNet fuses multi-channel complex spectra, GCC-PHAT spatial cues, and a face embedding at the bottleneck of a U-Net mask estimator, trained with auxiliary direction-of-arrival supervision and an SNR curriculum on simulated reverberant VoxCeleb mixtures. On a hard test set spanning $-1$ to 10~dB SNR, IsoNet reaches 9.31~dB SI-SDR, a 4.85~dB improvement over the mixture, and raises PESQ from 1.40 to 2.13 and STOI from 0.72 to 0.84. The gains are largest where the task is hardest (5.71~dB SI-SDR improvement at $-1$ to 1~dB SNR), and IsoNet outperforms oracle-steered delay-and-sum and MVDR beamformers by 9.7 and 10.9~dB SI-SDR. Ablations show that visual and spatial cues are complementary, and a 4.9M-parameter audio-spatial variant stays within 0.16~dB of the full model while running 26 times faster than real time.
\end{abstract}


\section{Introduction}

Human listeners can attend to one voice while suppressing competing talkers, an ability known as the cocktail party effect \cite{cherry1953cocktail}. They do so by combining binaural timing, room context, and the sight of the talker's face. Voice assistants, meeting devices, augmented reality (AR) headsets, hearing aids, and robots need the same ability, and the most natural interface is \emph{select-to-listen}: the user picks a visible person, and the device returns that person's voice. This is target speech extraction (TSE) \cite{zmolikova2023overview} with a visual cue, and it differs from blind separation because the output must match the person the user chose.

Compact devices make this hard. Classical beamformers such as delay-and-sum (DAS), minimum variance distortionless response (MVDR), and generalized sidelobe cancelers \cite{vanveen1988beamforming,capon1969mvdr,griffiths1982gsc,gannot2017consolidated} rely on an aperture that is large relative to the acoustic wavelength. A 9~cm array has weak angular resolution over most of the speech band, and reverberation further corrupts steering vectors and covariance estimates. Neural separators \cite{luo2019convtasnet,luo2020dualpath,subakan2021attention} learn strong speech priors but do not know which person the user selected, and most audio-visual systems \cite{ephrat2018looking,afouras2018conversation,gao2021visualvoice} are monaural or assume larger arrays \cite{gu2020multichannel}.

IsoNet closes this gap with a simple idea: let the face decide \emph{whom} to keep, and let learned spatial features recover the \emph{where} that the small array cannot resolve geometrically. The contributions are:
\begin{itemize}
    \item \textbf{A select-to-listen architecture.} IsoNet fuses 8-channel complex spectra, GCC-PHAT delay features, and a face embedding at the bottleneck of a U-Net mask estimator with auxiliary direction-of-arrival (DOA) supervision, so that a face selection directly steers extraction.
    \item \textbf{A compact-array audio-visual corpus.} A PyRoomAcoustics pipeline pairs VoxCeleb speech and face tracks with randomized reverberant rooms and a 9.4~cm tetrahedral array, providing clean targets, face crops, and source geometry for every mixture.
    \item \textbf{Large gains where beamforming fails.} IsoNet improves SI-SDR by 4.85~dB overall and by 5.71~dB at the lowest SNRs, and outperforms oracle-steered DAS and MVDR by 9.7 and 10.9~dB.
    \item \textbf{Ablation and efficiency analysis.} Face and spatial cues are complementary, extended GCC-PHAT lags add 0.75~dB, a moderate SNR curriculum improves every SNR bin, and a 4.9M-parameter variant runs 26 times faster than real time.
\end{itemize}

\section{Related work}

\subsection{Spatial filtering and multi-channel separation}

Microphone arrays enhance a desired direction by delaying and weighting channels before summation \cite{vanveen1988beamforming}. DAS is robust but has broad lobes on small arrays; MVDR \cite{capon1969mvdr} and generalized sidelobe cancelers \cite{griffiths1982gsc} reject interference more aggressively but depend on accurate steering vectors and covariance estimates \cite{gannot2017consolidated}. Mask-based neural beamformers \cite{heymann2016neural} estimate these statistics with a network, yet still rely on spatial differences between sources, which are small on compact arrays. GCC-PHAT \cite{knapp1976gcc} is the standard time-delay estimator, but in compact arrays the physically possible delays span only a few samples at 16~kHz, so spatial cues alone rarely separate nearby talkers.

\subsection{Neural separation and target speech extraction}

Supervised separation estimates time-frequency masks \cite{wang2014training} or time-domain representations \cite{luo2019convtasnet,luo2020dualpath,subakan2021attention}, using permutation-invariant training \cite{kolbaek2017pit} to resolve output order. U-Nets \cite{ronneberger2015unet} remain attractive for spectrogram-domain extraction because skip connections preserve harmonic detail while the bottleneck captures global context. TSE instead conditions the network on a cue for the desired talker, such as an enrollment utterance or a target direction \cite{zmolikova2023overview}. IsoNet uses a visual cue together with spatial features, so it needs neither enrollment audio nor a known DOA at inference.

\subsection{Audio-visual extraction}

Visual cues resolve which speaker to extract. Looking to Listen \cite{ephrat2018looking} conditions on face embeddings, The Conversation \cite{afouras2018conversation} uses lip motion, FaceFilter \cite{chung2020facefilter} uses a single still face image, and VisualVoice \cite{gao2021visualvoice} combines lip motion with cross-modal face identity; Michelsanti et al. \cite{michelsanti2021avse} give a broad survey. Gu et al. \cite{gu2020multichannel} showed that combining lip, voice, and directional cues improves multi-channel target separation, and active speaker detection \cite{tao2021talknet} shows the value of long-term visual dynamics.

Fusion can use concatenation, attention, or feature-wise modulation \cite{perez2018film}. IsoNet conditions the U-Net bottleneck with a compact MLP, which keeps the number of trainable parameters small and makes each modality's contribution measurable by ablation. Unlike the systems above, IsoNet combines a compact 4-microphone array, explicit GCC-PHAT spatial features, and user-selected face conditioning in a single model.

\section{Method}

\subsection{Signal model and extraction objective}

Let a compact array of $M$ microphones observe a reverberant mixture of target speech $s[n]$, interfering speech $u[n]$, room impulse responses $h_m^{(s)}$ and $h_m^{(u)}$, and sensor noise $v_m[n]$:
\begin{equation}
    x_m[n] = (s[n] * h_m^{(s)}[n]) + (u[n] * h_m^{(u)}[n]) + v_m[n].
    \label{eq:signal_model}
\end{equation}
In the STFT domain,
\begin{equation}
    X_m(f,t) \approx S(f,t)H_m^{(s)}(f,t) + U(f,t)H_m^{(u)}(f,t) + V_m(f,t).
    \label{eq:stft_model}
\end{equation}
The goal is to recover the target selected by the user through a face track. IsoNet estimates a magnitude mask $\hat{m}(f,t)$ for the reference microphone (channel 0) and reconstructs
\begin{equation}
    \hat{S}(f,t) = \hat{m}(f,t)|X_{ref}(f,t)|e^{j\angle X_{ref}(f,t)}.
    \label{eq:mask_reconstruction}
\end{equation}
Reference-phase reconstruction is stable and interpretable, which makes it well suited to isolating the effect of multimodal conditioning on mask estimation.

\subsection{Multi-channel spectral input}

For each microphone, IsoNet computes a complex STFT with a 512-point FFT, a Hann window, and a 10~ms hop. Real and imaginary parts from all microphones are stacked channel-wise:
\begin{equation}
    \mathbf{Z}(f,t)=\left[\Re(\mathbf{X}_1),\ldots,\Re(\mathbf{X}_M),\Im(\mathbf{X}_1),\ldots,\Im(\mathbf{X}_M)\right].
    \label{eq:complex_input}
\end{equation}
For $M=4$ this gives an 8-channel tensor that preserves inter-microphone phase differences, so the U-Net sees spatial information even before explicit GCC-PHAT features are added.

\subsection{GCC-PHAT spatial encoder}

For microphone pair $(i,j)$, the GCC-PHAT function \cite{knapp1976gcc} is
\begin{equation}
    R_{ij}(\Delta)=\mathcal{F}^{-1}\left\{\frac{X_i(f)X_j^*(f)}{|X_i(f)X_j^*(f)|+\epsilon}\right\}(\Delta).
    \label{eq:gcc_phat}
\end{equation}
The array yields six unique pairs; we keep 64 lags centered at zero, giving a $6 \times 64$ tensor. The maximum physical delay for the 9.4~cm aperture is only about 4.4 samples at 16~kHz, but the wider window also captures reflection side-lobes that carry room and position context for the learned encoder. A two-layer MLP maps the tensor to a 256-dimensional spatial embedding. An auxiliary head predicts $[\cos\phi,\sin\phi,\cos\theta,\sin\theta]$ of the target azimuth $\phi$ and elevation $\theta$, which avoids angle wraparound.

\subsection{Visual conditioning and U-Net fusion}

\begin{figure*}[!t]
    \centering
    \includegraphics[width=\textwidth]{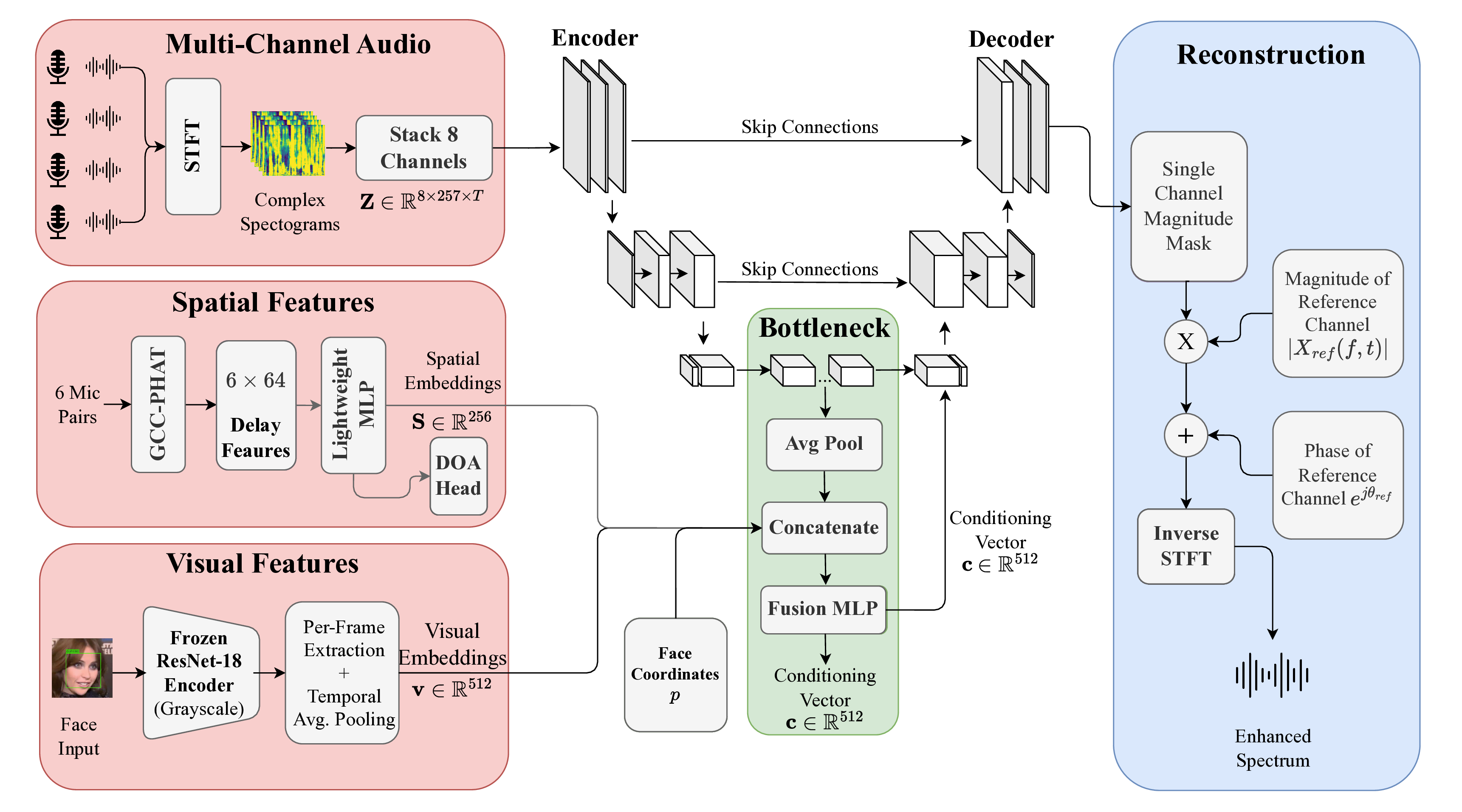}
    \caption{IsoNet architecture. Multi-channel complex spectra feed a U-Net; GCC-PHAT spatial and face embeddings condition its bottleneck to extract the user-selected speaker.}
    \label{fig:architecture}
\end{figure*}

The visual stream takes $112 \times 112$ grayscale face crops of the selected speaker. A frozen ResNet-18 \cite{he2016resnet} without its classification head extracts frame-level features, and temporal average pooling gives a 512-dimensional face embedding that encodes target identity and coarse visual context, similar in spirit to still-image conditioning \cite{chung2020facefilter}. Freezing the encoder reduces overfitting and training cost. The full architecture is shown in Fig.~\ref{fig:architecture}.

At the bottleneck, IsoNet pools the audio feature map to a 512-dimensional context vector $\mathbf{a}$ and concatenates it with the visual embedding $\mathbf{v}$, spatial embedding $\mathbf{s}$, and normalized face coordinates $\mathbf{p}$:
\begin{equation}
    \mathbf{c}=g([\mathbf{a},\mathbf{v},\mathbf{s},\mathbf{p}])\in\mathbb{R}^{512}.
    \label{eq:fusion}
\end{equation}
The fusion network $g(\cdot)$ is a two-layer MLP (input dimension 1282, ReLU, dropout 0.3). The conditioning vector is broadcast over the bottleneck feature map before decoding. The U-Net encoder has five stages with 32, 64, 128, 256, and 512 channels, and the decoder mirrors it with 256, 128, 64, and 32 channels before a single-channel mask output.

\subsection{Training objective and inference}

The model is trained with a magnitude reconstruction loss and an auxiliary DOA loss:
\begin{equation}
    \mathcal{L}=\left\lVert |\hat{\mathbf{S}}|-|\mathbf{S}|\right\rVert_1 + \lambda\left\lVert\hat{\mathbf{d}}-\mathbf{d}\right\rVert_2^2,\quad \lambda=0.5,
    \label{eq:loss}
\end{equation}
where $\mathbf{d}$ is the ground-truth angle vector and $\hat{\mathbf{d}}$ its prediction. We evaluate with the scale-invariant SDR \cite{leroux2019sdr}:
\begin{equation}
    \text{SI-SDR}(\mathbf{s},\hat{\mathbf{s}})=10\log_{10}\frac{\|\alpha\mathbf{s}\|^2}{\|\alpha\mathbf{s}-\hat{\mathbf{s}}\|^2},\quad \alpha=\frac{\langle\hat{\mathbf{s}},\mathbf{s}\rangle}{\|\mathbf{s}\|^2}.
    \label{eq:sisdr}
\end{equation}
At inference, IsoNet computes the STFTs and GCC-PHAT features, encodes the selected face crops, predicts the mask, and reconstructs the waveform by inverse STFT. The DOA head is used only during training; no oracle direction is needed at test time.

\section{Dataset and experimental setup}

\subsection{Simulation pipeline}

Aligned corpora with clean references, multi-channel audio, synchronized faces, and source geometry are scarce, so we build VoxCeleb-Sim from VoxCeleb \cite{nagrani2020voxceleb} using PyRoomAcoustics \cite{scheibler2018pyroomacoustics}. Each clip is trimmed to 4~s at 16~kHz, and per-frame face bounding boxes are stored and converted to grayscale crops.

\begin{figure}[!t]
    \centering
    \includegraphics[width=\columnwidth]{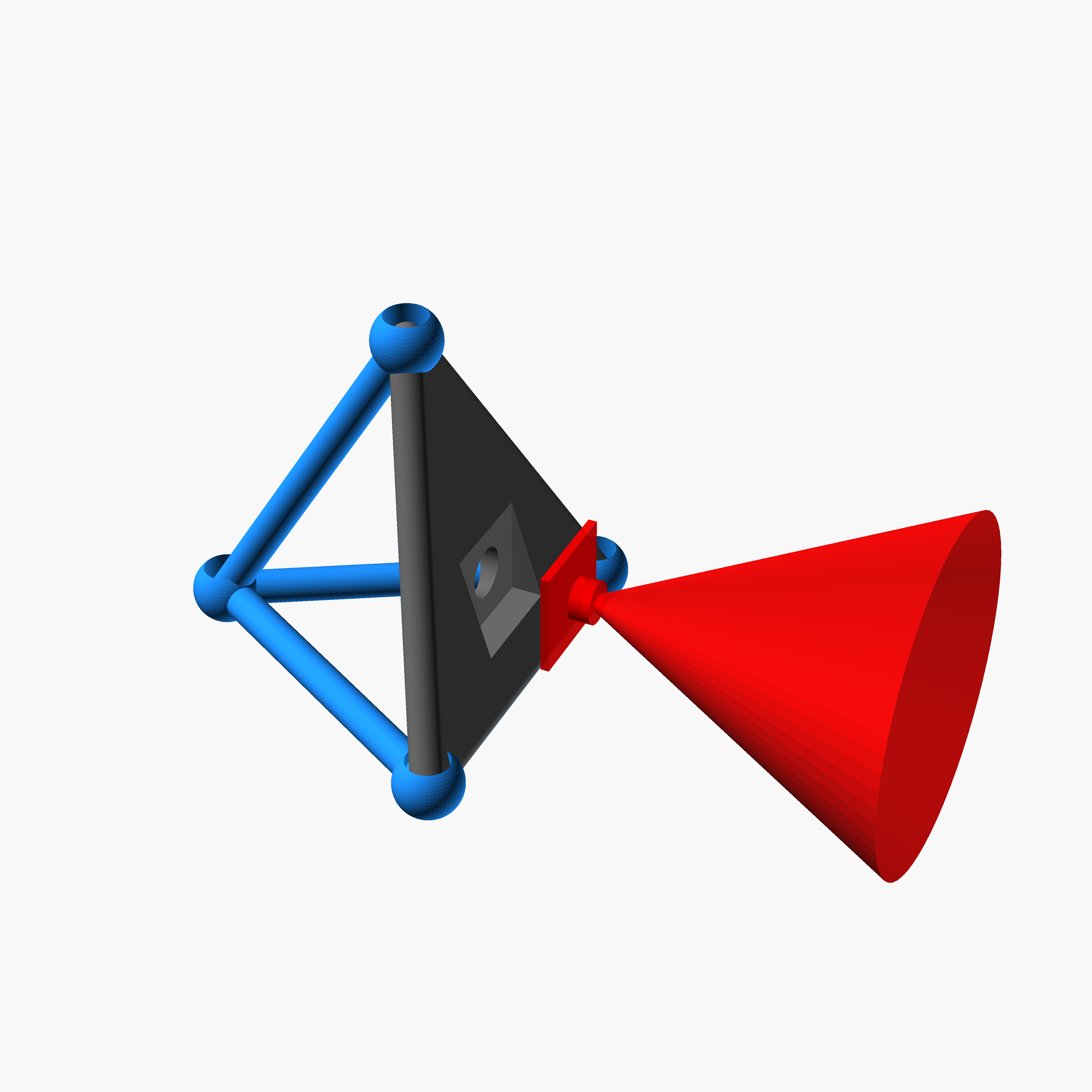}
    \caption{Compact tetrahedral 4-microphone array, camera, and visual field used in simulation.}
    \label{fig:array_geometry}
\end{figure}

The array is a compact tetrahedron: three microphones on a base circle of radius 5~cm and one 8~cm above its center (Fig.~\ref{fig:array_geometry}), giving a maximum baseline of about 9.4~cm, a size that fits small camera-equipped devices.

Rooms are randomized from $4\times3.5\times2.5$ to $10\times8\times3.5$~m with RT60 between 0.19 and 0.82~s. Targets lie in the camera field of view (azimuth within $\pm45^{\circ}$, elevation within $\pm20^{\circ}$, distance 0.8 to 1.5~m); interferers are placed randomly. Room impulse responses use the image source method \cite{allen1979image}, and sensor noise is added at $-50$~dB. Each sample contains the 4-channel mixture, the selected speaker's face frames, the clean target, the target location, and the room parameters (Table~\ref{tab:dataset_stats}).

\begin{table}[!t]
    \caption{VoxCeleb-Sim dataset statistics.}
    \label{tab:dataset_stats}
    \centering
    \small
    \begin{tabular}{@{}ll@{}}
        \toprule
        \textbf{Property}          & \textbf{Value}                    \\
        \midrule
        Samples                    & 25,000 clips                      \\
        Clip length                & 4.0 s                             \\
        Audio                      & 16 kHz, 4 channels                \\
        Train/test split           & 80/20                             \\
        SNR range                  & $-1$ to 20 dB (curriculum dependent) \\
        RT60                       & 0.19 to 0.82 s, mean 0.38 s       \\
        Target azimuth / elevation & $\pm45^{\circ}$ / $\pm20^{\circ}$ \\
        Target distance            & 0.8 to 1.5 m, mean 1.15 m         \\
        \bottomrule
    \end{tabular}
\end{table}

\subsection{Curriculum and optimization}

Following curriculum learning \cite{bengio2009curriculum}, we train three variants with increasingly difficult SNR ranges: IsoNet-Base on 5 to 20~dB, IsoNet-CL1 on 1 to 10~dB, and IsoNet-CL2 on $-1$ to 10~dB. All are evaluated on the same hard $-1$ to 10~dB test set, so the comparison measures generalization to difficult overlap.

All models train for 10 epochs with AdamW (learning rate $10^{-4}$, weight decay $10^{-5}$), cosine annealing, gradient clipping at 1.0, mixed precision, and an effective batch size of 16, taking 5 to 6 hours per model on an RTX 3090. The full model has 17.92M parameters, of which 6.75M are trainable and 11.17M belong to the frozen visual encoder.

\subsection{Metrics}

We report SI-SDR \cite{leroux2019sdr} as the primary metric, its improvement over the mixture (SI-SDRi), SDR \cite{vincent2006performance}, PESQ \cite{rix2001pesq} for perceptual quality, and STOI \cite{taal2011stoi} for intelligibility.

\section{Results}

\begin{figure*}[!t]
    \centering
    \includegraphics[width=\textwidth]{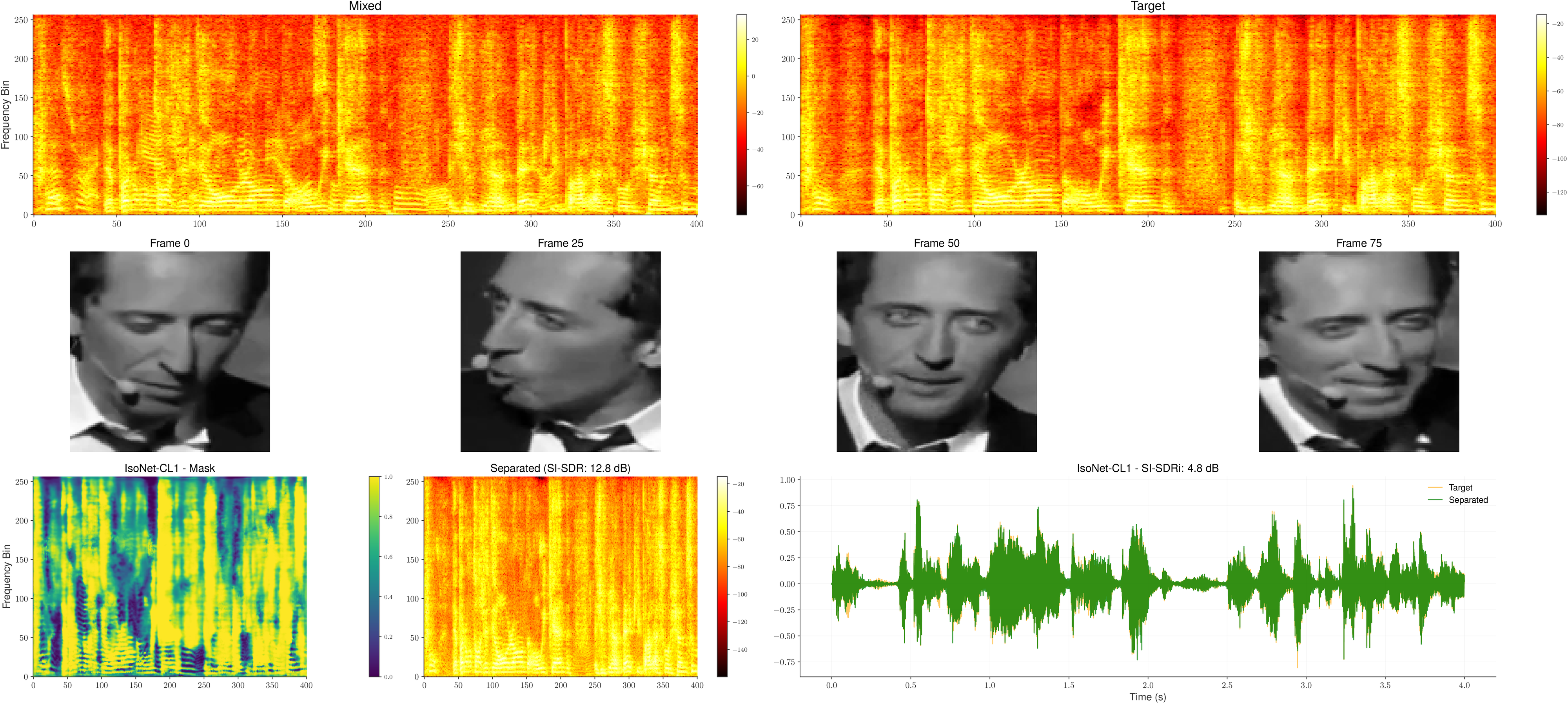}
    \caption{Representative test sample. Top: mixture and clean target spectrograms. Middle: frames from the selected speaker's face track. Bottom: IsoNet-CL1 mask, separated spectrogram, and separated waveform overlaid on the target.}
    \label{fig:qualitative}
\end{figure*}

\subsection{Main results}

Table~\ref{tab:main_results} summarizes performance on the hard test set. Every IsoNet variant improves every metric by a wide margin. The best model, IsoNet-CL1, reaches 9.31~dB SI-SDR, a 4.85~dB improvement, raises SDR by 5.07~dB, lifts PESQ from 1.40 to 2.13 (+52\%), and lifts STOI from 0.72 to 0.84. Moderate curriculum hardening gives the best model; extending the training range down to $-1$~dB (CL2) brings no further gain, so CL1 is our default.

\begin{table}[!t]
    \caption{Results on the hard $-1$ to 10~dB test set (means over 4,369 mixtures; SI-SDR, SI-SDRi, and SDR in dB). Beamformers use the oracle target direction; IsoNet does not.}
    \label{tab:main_results}
    \centering
    \small
    \adjustbox{max width=\columnwidth}{
        \begin{tabular}{@{}lccccc@{}}
            \toprule
            \textbf{Method} & \textbf{SI-SDR} & \textbf{SI-SDRi} & \textbf{SDR}  & \textbf{PESQ} & \textbf{STOI} \\
            \midrule
            Mixture         & 4.46            & --               & 4.51          & 1.40          & 0.72          \\
            DAS             & $-0.36$         & $-4.82$          & --            & 1.35          & 0.65          \\
            MVDR            & $-1.62$         & $-6.08$          & --            & 1.27          & 0.64          \\
            \midrule
            IsoNet-Base     & 8.62            & 4.16             & 8.78          & 1.98          & 0.83          \\
            IsoNet-CL1      & \textbf{9.31}   & \textbf{4.85}    & \textbf{9.58} & \textbf{2.13} & \textbf{0.84} \\
            IsoNet-CL2      & 9.13            & 4.68             & 9.36          & 2.11          & \textbf{0.84} \\
            \bottomrule
        \end{tabular}}
\end{table}

\textbf{Beamforming comparison.} With a 9.4~cm aperture, oracle-steered DAS and MVDR have main lobes too broad to separate two talkers across most of the speech band, and MVDR covariance estimates from 4~s reverberant clips are unreliable; both fall below the unprocessed mixture. IsoNet uses the same four channels without oracle direction and outperforms DAS by 9.67~dB and MVDR by 10.93~dB SI-SDR. On compact arrays, target extraction therefore calls for learned multimodal conditioning rather than geometric steering alone.

\subsection{Robustness across SNR}

Table~\ref{tab:per_snr} stratifies the test set by input SNR. IsoNet helps most where help is needed most: SI-SDRi rises steadily as the input gets harder, from 3.96~dB in the [7,10]~dB bin to 5.71~dB in the [$-1$,1)~dB bin, turning a 0~dB input into a 5.7~dB output. Curriculum training improves every bin over IsoNet-Base, again with the largest margin (+0.86~dB) in the hardest bin.

\begin{table}[!t]
    \caption{SI-SDRi (dB) by input SNR on the hard test set. Input: SI-SDR of the unprocessed mixture.}
    \label{tab:per_snr}
    \centering
    \small
    \begin{tabular}{@{}lccccc@{}}
        \toprule
        \textbf{SNR (dB)} & \textbf{N} & \textbf{Input} & \textbf{Base} & \textbf{CL1}  & \textbf{CL2} \\
        \midrule
        $[-1,1)$          & 765        & 0.0            & 4.85          & \textbf{5.71} & 5.40         \\
        $[1,3)$           & 843        & 2.0            & 4.64          & \textbf{5.42} & 5.19         \\
        $[3,5)$           & 801        & 4.0            & 4.25          & \textbf{4.97} & 4.77         \\
        $[5,7)$           & 809        & 6.0            & 3.96          & \textbf{4.60} & 4.48         \\
        $[7,10]$          & 1151       & 8.4            & 3.44          & \textbf{3.96} & 3.90         \\
        \bottomrule
    \end{tabular}
\end{table}

\subsection{Ablation studies}

\textbf{Modalities.} Table~\ref{tab:modality_ablation} isolates each conditioning stream. The audio-only U-Net is already strong (8.87~dB) because its 8-channel complex input carries inter-channel phase. Adding the face embedding (+0.30~dB) or the GCC-PHAT embedding (+0.28~dB) each improves all metrics, and combining them is best on every metric (+0.44~dB SI-SDR, +0.13 PESQ, +0.015 STOI). The cues are complementary: identity tells the network \emph{whom} to keep, and delay structure tells it \emph{where} that person is. The spatial branch alone delivers nearly two-thirds of the full gain for only 0.72M extra parameters.

\begin{table}[!t]
    \caption{Modality ablation on the hard test set (A: audio, V: visual, S: spatial).}
    \label{tab:modality_ablation}
    \centering
    \small
    \begin{tabular}{@{}lcccc@{}}
        \toprule
        \textbf{Variant} & \textbf{Params (M)} & \textbf{SI-SDR (dB)} & \textbf{PESQ} & \textbf{STOI}  \\
        \midrule
        Audio-only       & 4.19                & 8.87                 & 2.00          & 0.826          \\
        A+V              & 17.46               & 9.17                 & 2.06          & 0.838          \\
        A+S              & 4.91                & 9.15                 & 2.06          & 0.833          \\
        A+V+S (full)     & 17.92               & \textbf{9.31}        & \textbf{2.13} & \textbf{0.841} \\
        \bottomrule
    \end{tabular}
\end{table}

\textbf{GCC-PHAT lags.} Widening the GCC-PHAT window from 16 to 64 lags raises SI-SDR from 8.56 to 9.31~dB, PESQ from 1.94 to 2.13, and STOI from 0.823 to 0.841. Since 16 lags already cover the $\pm$4.4-sample direct-path range, the gain comes from reflection structure beyond the physical delay range, which the learned encoder turns into useful room and position context.

\subsection{Efficiency}

Table~\ref{tab:compute_cost} reports cost per 4~s clip on an entry-level RTX 3050 GPU. All variants run faster than real time. The audio-spatial variant needs only 4.91M parameters and under 9 GFLOPs and runs at a real-time factor of 0.039, about 26 times faster than real time, while staying within 0.16~dB of the full model. About 95\% of the full model's FLOPs come from applying the frozen ResNet-18 to every frame; because the face embedding is temporally averaged, it can be computed on sparse frames or cached per tracked face, bringing the full model's cost close to that of the audio-spatial variant.

\begin{table}[!t]
    \caption{Cost per 4~s clip on an RTX 3050 GPU (RTF: real-time factor).}
    \label{tab:compute_cost}
    \centering
    \small
    \begin{tabular}{@{}lcccc@{}}
        \toprule
        \textbf{Variant} & \textbf{Params (M)} & \textbf{GFLOPs} & \textbf{Time (ms)} & \textbf{RTF} \\
        \midrule
        Audio-only       & 4.19                & 8.7             & 145.5              & 0.036        \\
        A+S              & 4.91                & 8.9             & 154.9              & 0.039        \\
        A+V+S (full)     & 17.92               & 189.1           & 596.9              & 0.149        \\
        \bottomrule
    \end{tabular}
\end{table}

\subsection{Qualitative analysis}

\begin{figure}[!t]
    \centering
    \includegraphics[width=0.85\columnwidth]{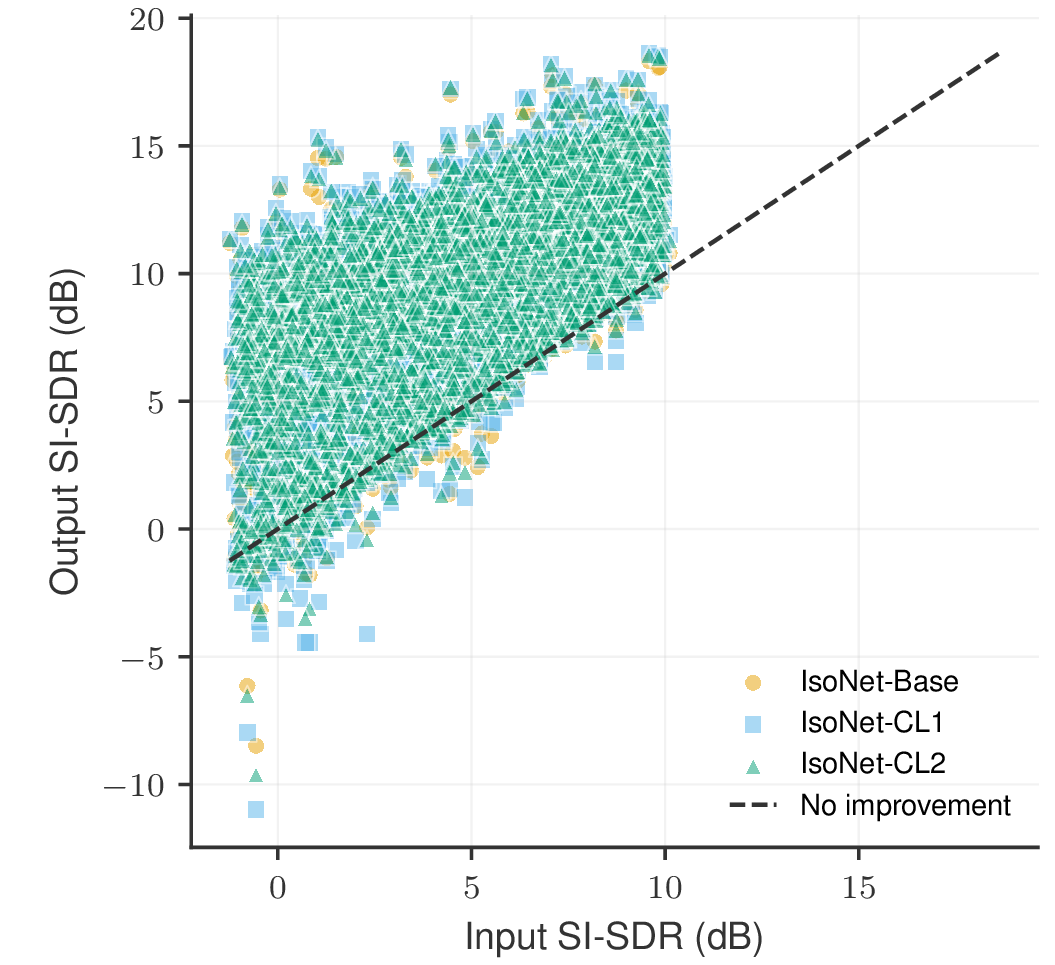}
    \caption{Output versus input SI-SDR for every test mixture. Points above the dashed line are improved by IsoNet.}
    \label{fig:scatter}
\end{figure}

Fig.~\ref{fig:qualitative} shows a representative example: the IsoNet-CL1 mask removes most of the interfering talker's energy, and the separated waveform closely tracks the clean target. Fig.~\ref{fig:scatter} plots output against input SI-SDR for every test mixture. The large majority of points lie above the no-improvement line, so the gains are consistent across the test set rather than driven by a few samples.

\subsection{Limitations and outlook}
\label{sec:limitations}

This study uses simulated rooms with one interfering talker. Real deployment adds camera--array calibration and audio-video synchronization, and evaluation on real recordings is the natural next step. Further extensions include complex masking in place of mixture phase, temporal lip encoders, and scenes with several interferers.

\section{Conclusion}

IsoNet lets a user select a face and hear that person's voice from a 9.4~cm, 4-microphone array. By conditioning a U-Net mask estimator on face identity and GCC-PHAT spatial cues, it improves SI-SDR by 4.85~dB on a hard $-1$ to 10~dB test set (5.71~dB at the lowest SNRs), raises PESQ by 0.73 and STOI by 0.12, and outperforms oracle-steered DAS and MVDR beamformers by more than 9.6~dB. Visual and spatial cues contribute complementary gains, and a lightweight audio-spatial variant runs 26 times faster than real time. Compact camera-equipped devices can therefore listen selectively in a regime where classical spatial filtering does not work.

\section*{Acknowledgements}

This research did not receive any specific grant from funding agencies in the public, commercial, or not-for-profit sectors.

\section*{Conflict of interest}

The authors declare that they have no conflict of interest.

\section*{Data availability}

The data supporting the findings of this study are available from the corresponding author upon reasonable request.

\bibliographystyle{IEEEtran}
\bibliography{ref}

\end{document}